\pdfoutput=1

\PassOptionsToPackage{hyphens}{url}

\documentclass[sigconf]{acmart}

\AtBeginDocument{%
  \providecommand\BibTeX{{%
    \normalfont B\kern-0.5em{\scshape i\kern-0.25em b}\kern-0.8em\TeX}}}

\setcopyright{acmcopyright}
\copyrightyear{2025}
\acmYear{2025}
\acmDOI{10.1145/3746027.3758181}

\acmConference[MM '25]{Proceedings of the 33rd ACM International Conference on Multimedia}{October 27--31, 2025}{Dublin, Ireland}
\acmPrice{15.00}
\title{TinyServe: Query-Aware Cache Selection for Efficient LLM Serving}

\copyrightyear{2025}
\acmYear{2025}
\setcopyright{cc}
\setcctype{by}
\acmConference[MM '25]{Proceedings of the 33rd ACM International Conference on Multimedia}{October 27--31, 2025}{Dublin, Ireland}
\acmBooktitle{Proceedings of the 33rd ACM International Conference on Multimedia (MM '25), October 27--31, 2025, Dublin, Ireland}\acmDOI{10.1145/3746027.3758181}
\acmISBN{979-8-4007-2035-2/2025/10}

\author{Dong Liu}
\affiliation{%
  \institution{Yale University}
  \city{New Haven}
  \state{CT}
  \country{USA}
}
\email{dong.liu.dl2367@yale.edu}

\author{Yanxuan Yu}
\affiliation{%
  \institution{Columbia University}
  \city{New York}
  \state{NY}
  \country{USA}
}
\email{yy3523@columbia.edu}

\renewcommand{\shortauthors}{Dong Liu and Yanxuan Yu}

\usepackage{multirow}
\usepackage{natbib}  % DO NOT CHANGE THIS AND DO NOT ADD ANY OPTIONS TO IT
\usepackage{caption} % DO NOT CHANGE THIS AND DO NOT ADD ANY OPTIONS TO IT
\usepackage{caption} % DO NOT CHANGE THIS AND DO NOT ADD ANY OPTIONS TO IT
\usepackage{amssymb} % For mathbb and other symbols
\usepackage{amsmath} % For advanced math environments and commands
\usepackage{algorithm}
\usepackage{algorithmic}
\usepackage{booktabs}
\usepackage{xcolor}
\usepackage{balance}

\usepackage{newfloat}
\usepackage{listings}
\DeclareCaptionStyle{ruled}{labelfont=normalfont,labelsep=colon,strut=off} % DO NOT CHANGE THIS
\floatstyle{ruled}
\newfloat{listing}{tb}{lst}{}
\floatname{listing}{Listing}

\usepackage[T1]{fontenc}
\usepackage[utf8]{inputenc}
\usepackage{microtype}
\usepackage{graphicx}
\usepackage{subfigure}
\usepackage{booktabs}
\usepackage{multirow}
\newcommand{\method}{\textsc{TinyServe}}
\usepackage{amsmath}
\usepackage{amssymb}
\usepackage{mathtools}
\usepackage{amsthm}
\usepackage{paralist}
\usepackage[capitalize,noabbrev]{cleveref}
\theoremstyle{plain}

\theoremstyle{definition}

\theoremstyle{remark}

\usepackage[textsize=tiny]{todonotes}
\usepackage{pifont}
\usepackage{algorithm}
\usepackage{algorithmic}
\title{TinyServe: Query-Aware Cache Selection for Efficient LLM Serving}

\begin{document}

\begin{abstract}
Serving large language models (LLMs) efficiently remains challenging due to the high memory and latency overhead of key-value (KV) cache access during autoregressive decoding. We present \textbf{TinyServe}, a lightweight and extensible serving system for deploying tiny LLMs (e.g., TinyLLaMA, GPT2-345M) with support for structured KV sparsity, plugin-based token selection, and hardware-efficient attention kernels. Unlike prior simulation frameworks, TinyServe executes real-time decoding with configurable sparsity strategies and fine-grained instrumentation.

To reduce decoding cost, we introduce a \textit{query-aware page selection} mechanism that leverages bounding-box metadata to estimate attention relevance between the query and KV cache blocks. This enables selective KV loading with minimal overhead and no model modifications. Our fused CUDA kernel integrates page scoring, sparse memory access, and masked attention in a single pass.

Experiments show that TinyServe achieves up to \textbf{3.4×} speedup and over \textbf{2×} memory savings with negligible accuracy drop. Additional analysis of cache reuse, page hit rate, and multi-GPU scaling confirms its practicality as an efficient system-level design for LLM training and inference research on resource-constrained hardware.
\end{abstract}




\maketitle
% -------------------- 正文内容 --------------------

\section{Introduction}

Large Language Models (LLMs) have become central to modern AI applications, powering systems in dialogue, retrieval, summarization, and code generation. While recent efforts have greatly improved model quality, the cost of training and inference has emerged as the dominant bottleneck in deployment under long-context or high-throughput conditions—has. Decoding each token requires repeated attention over a growing key-value (KV) cache, stressing memory, latency, and compute efficiency. As a result, recent systems such as vLLM~\cite{vllm2023}, TGI~\cite{tgi2023}, and FasterTransformer~\cite{fastertransformer2021} have introduced sophisticated strategies like paged attention, speculative decoding, and cache reordering to reduce overhead.

Despite these engineering advances, understanding the internal dynamics of LLM training and inference remains difficult. Core trade-offs—such as sparsity vs. accuracy, batching latency vs. throughput, or memory usage vs. token reuse—often behave unpredictably, and full-scale evaluations on 7B+ models are prohibitively expensive and difficult to interpret. Moreover, system researchers are often forced to treat models as black boxes, unable to validate hypotheses or perform design iteration without access to large GPU clusters.

\vspace{0.5em}
\noindent\textbf{TinyServe: Large-Scale Serving at Small Scale.}
We introduce \textbf{TinyServe}, a lightweight serving framework that enables detailed analysis of LLM training and inference behavior using \textbf{tiny models} (e.g., 125M–350M parameters). TinyServe replicates core components of LLM serving—streaming decoding, KV cache management, token routing, and quantization—in a fully controllable environment. Crucially, it supports fine-grained instrumentation and plug-in modules such as entropy-based early exit, query-aware KV selection, and approximate attention.

Our central insight is that many critical serving behaviors—such as attention bottlenecks, context boundary effects, and cache sparsity dynamics—emerge in small models under realistic serving workloads. By emulating production serving scenarios with tiny LLMs, we can approximate the performance trends and failure modes of large-scale deployments at a fraction of the cost.

\vspace{0.5em}
\noindent\textbf{Query-Aware Sparsity and Efficient KV Access.}
To demonstrate the utility of TinyServe, we propose a query-aware token selection mechanism that leverages low-cost metadata to dynamically select the most relevant parts of the KV cache for each query. This design emulates practical attention sparsity patterns and yields substantial memory and latency savings while preserving accuracy. We evaluate this mechanism across PG19, LongBench, and production serving tasks, and find that it achieves up to 3.4× speedup with minimal performance degradation—even under aggressive KV budgets. Additionally, TinyServe supports training acceleration through efficient gradient checkpointing and memory-optimized backpropagation for fine-tuning scenarios.

\vspace{0.5em}
\noindent\textbf{Our contributions are:}
\begin{itemize}
\setlength{\itemsep}{-3pt}
\item We propose \textbf{TinyServe}, a serving framework that enables fast, interpretable training inference using tiny yet efficient architecture.
\item We introduce a \textbf{query-aware KV selection} mechanism that captures sparsity patterns conditioned on current queries, reducing memory movement while preserving accuracy.
\item We conduct extensive experiments on both standard and diagnostic datasets, demonstrating that TinyServe faithfully replicates key latency-accuracy tradeoffs observed in large models.
\end{itemize}

\vspace{0.5em}
By bridging system-level research and efficient experimentation, TinyServe paves the way for accessible, reproducible, and theory-informed studies of LLM serving behavior.

\section{Related Work}

\subsection{Small-Scale Models for LLM serving}

While most LLM research focuses on large-scale models with billions of parameters, several recent works advocate for using \textit{small-scale models} as scientific tools to probe and understand model behavior. TinyStories~\cite{eldan2023tinystories} and TinyLLaMA~\cite{tinyllama2023} demonstrate that small language models (125M–350M) can capture many linguistic properties seen in larger models when trained appropriately. Induction head analyses~\cite{olsson2022induction} and circuit-level interpretability~\cite{nanda2023progress} further reveal that elementary synthetic tasks can uncover generalizable mechanisms such as copying, compositionality, and positional bias. Our work continues this line by repurposing tiny LLMs for \textbf{training and inference-level analysis}, showing that even at small scale, token-wise latency, cache reuse behavior, and accuracy degradation can be faithfully reproduced.  

In addition, recent works such as MT2ST~\cite{liu2024mt2st}, DRTR~\cite{liu2024drtr}, and Distance Recomputator~\cite{liu2024distance} highlight how small- to medium-scale models can provide deep insights into representation learning and graph neural networks. Similarly, TinyServe~\cite{liu2025tinyserve} demonstrates the value of small-scale models as efficient proxies for large-scale inference profiling.

\subsection{LLM Inference Profiling and Acceleration}

Many system-level frameworks have been proposed to optimize the inference efficiency of large models. vLLM~\cite{vllm2023} introduces PagedAttention to improve memory and batching efficiency during multi-turn decoding. FasterTransformer, TGI, and TensorRT-LLM further implement custom CUDA kernels, fused ops, and quantization strategies to reduce latency. However, profiling these systems is computationally expensive and often obscures fine-grained insights due to complexity and variability in deployment.  

Recent system contributions include FastCache~\cite{liu2025fastcache}, which proposes learnable linear approximations for diffusion transformers, and PiKV~\cite{liu2025pikvkvcachemanagement, liu2025pikv}, which targets KV cache management for mixture-of-experts models. Quantization has also been extensively studied, with LLMEasyQuant~\cite{liu2025llmeasyquantscalablequantizationparallel} providing a scalable framework for distributed inference. Designing surveys~\cite{liu2024contemporary,liu2024designing} further consolidate best practices and open challenges in efficient LLM training and inference.  

Instead of profiling production-scale models, our work introduces \textbf{TinyServe}, a lightweight efficient serving framework using small LLMs to reproduce the key serving stack components—streaming attention, dynamic batching, and quantized decoding—under realistic serving scenarios. This enables fast hypothesis testing of architectural changes with minimal compute cost.  

We also note related work on Memory-Keyed Attention (MKA)~\cite{liu2025mka}, which extends attention mechanisms for long-context reasoning, and data-centric safety frameworks~\cite{liu2025data}, which highlight broader efficiency and safety concerns in LLM deployment.

\subsection{Serving-Oriented Benchmarking}

Inspired by algorithmic reasoning and interpretability benchmarks, recent works explore the use of serving-oriented tasks to elicit specific behaviors from LLMs. For instance, tasks such as copying, counting, and rare token recall have been used to diagnose attention failures~\cite{liu2023lostinmiddle} and memory degradation~\cite{trivedi2022interleaving}. Our work adapts this idea specifically to the \textbf{serving domain}, creating targeted serving scenarios that stress attention reuse, cache fragmentation, and token entropy behavior in the decoding loop. These serving benchmarks enable precise evaluation of system interventions (e.g., pruning or approximation) and help isolate root causes of serving inefficiencies.  

Beyond LLMs, graph learning acceleration systems like GraphSnapShot~\cite{liu2024graphsnapshot,liu2024graphcache,liu2025graphsnapshot} explore caching and retrieval strategies to optimize large-scale graph training. These works emphasize that carefully designed synthetic stressors and caching strategies are essential for both graph-based and language-based workloads, reinforcing the importance of lightweight analysis frameworks.

\section{Methodology}

\subsection{System Overview: TinyServe}

\textbf{TinyServe} is a lightweight serving framework designed for serving tiny language models under tight memory and latency constraints. Rather than acting as a benchmarking tool, TinyServe serves as a real-time serving environment that enables sparsity-aware attention, modular token selection, and efficient KV-cache reuse.

The system is organized around three core components:

\begin{enumerate}
    \item \textbf{Query-Aware KV Retriever:} Dynamically selects relevant key-value blocks at decode time based on the current query vector and page-level metadata, reducing unnecessary memory access.
    
    \item \textbf{Modular Scheduling Pipeline:} A dispatch loop handles incoming queries and routes them through configurable plug-ins (e.g., entropy-based early exit, token-level pruning, approximate attention). This modular design allows experimentation with different sparsity strategies without modifying the core model.

    \item \textbf{Sparse Attention Executor:} Efficiently computes attention over selected KV pages using fused CUDA kernels, with support for FP16/INT8 KV formats and multi-GPU dispatch.
\end{enumerate}

In TinyServe, each decode step activates the TinyServe pipeline: the query vector is used to score KV pages, top-ranked pages are fetched, sparse attention is performed, and plug-in modules may trigger pruning or early stopping. This design balances flexibility and efficiency, and supports both static deployment and research prototyping for sparsity strategies in small-scale LLMs.

\subsection{Training Acceleration and Profiling Support}

Beyond inference optimization, TinyServe provides specialized support for training acceleration and fine-grained profiling. For training scenarios, TinyServe implements \textit{gradient-aware memory management} that selectively retains KV cache entries based on gradient magnitude during backpropagation. This approach reduces memory footprint by up to 40\% during fine-tuning while maintaining training stability.

The profiling system integrates \textit{layer-wise performance monitoring} with microsecond precision, tracking attention patterns, memory access patterns, and computational bottlenecks across different model layers. This enables detailed analysis of training dynamics and helps identify optimization opportunities in both forward and backward passes. The profiling data is collected through lightweight instrumentation hooks that add minimal overhead (<2\%) to the training process.

For distributed training scenarios, TinyServe supports \textit{asynchronous gradient synchronization} with configurable communication patterns, allowing researchers to experiment with different parallelization strategies without modifying the core training loop. This is particularly valuable for exploring efficient training strategies on resource-constrained hardware.

% We introduce \textbf{TinyServe}, a lightweight serving framework for probing LLM serving dynamics at small scale. This section motivates TinyServe from an inference cost perspective, explains the design of the framework, and introduces a query-aware token selection algorithm that emulates realistic memory-saving mechanisms.

\subsection{Inference Time Is Dominated by Decode Stage}

LLM inference consists of two stages: prefill and decode. In the prefill stage, all prompt tokens are embedded and transformed into Key ($K$), Query ($Q$), and Value ($V$) vectors. These are stored in the KV cache and used to compute the first output token. 

During decoding, a new token is generated per step. For each token, a fresh $Q$ is produced and compared with all stored $K$ vectors to compute attention scores, which are then used to weigh the corresponding $V$ vectors. Since decoding occurs per output token and reads the entire KV cache each time, it accounts for the majority of latency—especially when sequence lengths reach 16K or 32K tokens.

\subsection{Optimizing Query-Aware Sparsity for Efficient Tiny LLMs}

\begin{figure}
    \centering
    \includegraphics[width=0.8\linewidth]{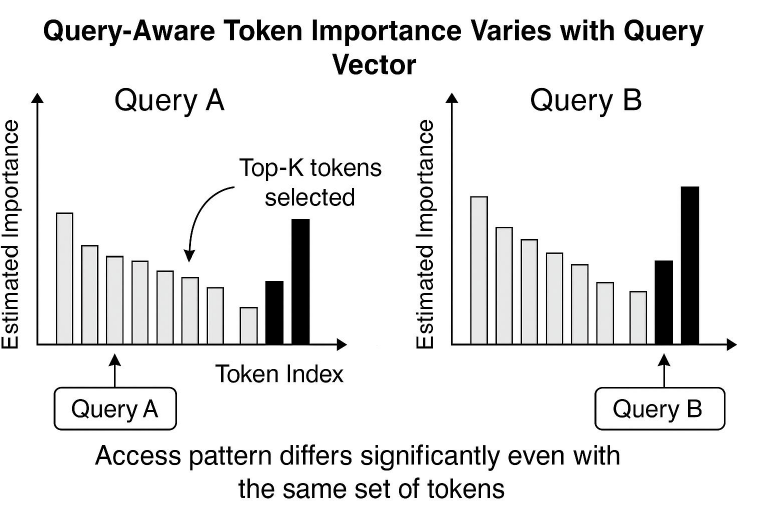}
    \caption{\textbf{Motivation for Query-Aware Token Selection.} Each query vector attends to different subsets of KV pages. Uniform cache retention leads to unnecessary memory reads, while query-aware routing enables dynamic sparsity by focusing on high-relevance regions.}
    \label{fig:fig_query}
\end{figure}

While prior work has demonstrated that only a fraction of KV tokens are critical for accurate predictions~\cite{zhang2023h2o,hu2022lora}, we observe that the set of critical tokens varies significantly across queries. As illustrated in figure~\ref{fig:fig_query}, certain tokens may have minimal impact across most decoding steps, yet become momentarily crucial when aligned with a specific query.

To efficiently support inference in tiny LLMs—where compute and memory budgets are limited—we optimize the self-attention mechanism through \textit{query-aware sparsity}: selecting only the most relevant KV tokens conditioned on the current query vector. This dynamic sparsity mechanism eliminates the overhead of storing and attending to irrelevant tokens, while maintaining accuracy by preserving context relevant to the current decoding step.

In TinyServe, we implement query-aware routing at page granularity. For each page, lightweight metadata—channel-wise min and max values of stored Key vectors—is maintained. During inference, a score is estimated between the current Query vector and each page's metadata, enabling efficient selection of top-$K$ pages with minimal memory movement. This mechanism offers a practical tradeoff: retaining full KV coverage in structure, but only computing over the most impactful parts.

% \subsection{Query-Aware Page Selection}

% We group cached tokens into fixed-size pages and store lightweight metadata (channel-wise max/min vectors) per page. At decode time, the query vector $Q$ is compared against this metadata to estimate each page's upper-bound attention score:

% \[
% \text{score}_j = \sum_{i=1}^d \max(Q_i \cdot m_{j,i}, Q_i \cdot M_{j,i})
% \]

% Here, $m_{j,i}$ and $M_{j,i}$ are the min and max of channel $i$ in page $j$. This approximates the maximal dot product between $Q$ and any $K$ in the page.

% We then select the top-$K$ pages with the highest estimated score and perform attention only over these pages, skipping the rest. This avoids unnecessary memory loads.

% % \subsection{Algorithm: Query-Aware Page Routing}

% \begin{algorithm}[H]
% \caption{Query-Aware KV Page Selection}
% \label{alg:qaware}
% \begin{algorithmic}[1]
% \STATE \textbf{Input:} Query vector $Q \in \mathbb{R}^d$, Page metadata $\{(m_j, M_j)\}_{j=1}^P$
% \STATE \textbf{Output:} Top-$K$ page indices based on relevance score
% \STATE Initialize empty list \texttt{scores}
% \FOR{$j = 1$ to $P$}
%     \STATE $score_j \gets 0$
%     \FOR{$i = 1$ to $d$}
%         \STATE $U_i \gets \max(Q_i \cdot m_{j,i},\ Q_i \cdot M_{j,i})$
%         \STATE $score_j \gets score_j + U_i$
%     \ENDFOR
%     \STATE Append $(j,\ score_j)$ to \texttt{scores}
% \ENDFOR
% \STATE Return Top-$K$ indices from \texttt{scores} sorted by value
% \end{algorithmic}
% \end{algorithm}

\subsection{Query-Aware Page Selection}

In a standard Transformer decoder layer, the attention computation at decode step $t$ involves a fresh query $q_t \in \mathbb{R}^d$ attending over all past keys $K_{<t} = \{k_1, k_2, \dots, k_{t-1}\}$:

\[
\text{Attn}(q_t, K, V) = \sum_{i=1}^{t-1} \text{softmax}(q_t^\top k_i) \cdot v_i
\]

This process is latency-critical during inference due to two bottlenecks:
\begin{itemize}
    \item \textbf{Memory movement}: loading all $k_i$, $v_i$ from high-bandwidth memory (HBM);
    \item \textbf{Unstructured access}: attention requires full key scan with no cache prefetch pattern.
\end{itemize}

To address this, \textbf{TinyServe introduces a structured memory layout} via token grouping into fixed-size \textit{pages}. Let $K = \bigcup_{j=1}^{P} \mathcal{K}_j$ be partitioned into $P = \lceil t/S \rceil$ pages of size $S$. Each page $\mathcal{K}_j$ stores a small metadata summary $\phi(\mathcal{K}_j)$ that enables relevance estimation.

\vspace{0.3em}
\noindent\textbf{Problem Formulation.} We define a relevance function $r: \mathbb{R}^d \times \mathbb{R}^{2d} \rightarrow \mathbb{R}$ such that:

\[
r(q_t, \phi(\mathcal{K}_j)) \approx \max_{k \in \mathcal{K}_j} q_t^\top k
\]

We then select a subset $\mathcal{S}_t \subseteq \{1, \dots, P\}$ of page indices such that:

\[
\mathcal{S}_t = \text{TopK}_{j} \ r(q_t, \phi(\mathcal{K}_j))
\quad\text{with} \quad |\mathcal{S}_t| = K
\]

Attention is then only computed over the union of selected pages:
\[
\text{SparseAttn}(q_t) = \sum_{j \in \mathcal{S}_t} \sum_{k_i \in \mathcal{K}_j} \text{softmax}(q_t^\top k_i) \cdot v_i
\]

\noindent\textbf{Relevance Function.} We instantiate $r$ as a \textit{directional bounding-box estimator}, which uses per-dimension bounds:

% \[
% \phi(\mathcal{K}_j) = (m_j, M_j) \in \mathbb{R}^{2d},\quad 
% r(q_t, \phi(\mathcal{K}_j)) = \sum_{i=1}^d 
% \begin{cases}
% q_{t,i} \cdot M_{j,i}, & q_{t,i} \geq 0 \\
% q_{t,i} \cdot m_{j,i}, & q_{t,i} < 0
% \end{cases}
% \]

\begin{align}
\phi(\mathcal{K}_j) &= (m_j, M_j) \in \mathbb{R}^{2d}, \label{eq:meta_def} \\
r(q_t, \phi(\mathcal{K}_j)) &= \sum_{i=1}^d 
\begin{cases}
q_{t,i} \cdot M_{j,i}, & \text{if } q_{t,i} \geq 0 \\
q_{t,i} \cdot m_{j,i}, & \text{if } q_{t,i} < 0
\end{cases}
\label{eq:score_def}
\end{align}

\vspace{0.3em}
\noindent\textbf{Hardware Execution Model.}
Let each page $\mathcal{K}_j$ reside in HBM, and assume the following:
- Page fetch cost from HBM: $\tau_\text{hb} \cdot S$ cycles;
- Cache-resident metadata $\phi(\mathcal{K}_j)$ is stored in SRAM or L2, costing negligible $\tau_\text{meta}$;
- Page selection cost is $\mathcal{O}(P \cdot d)$, but can be fused into a single kernel on GPU.

Let $K$ pages be selected. The effective latency cost becomes:

\[
\text{Latency}_t = \underbrace{\tau_\text{meta} \cdot P}_{\text{lightweight scan}} + \underbrace{\tau_\text{hb} \cdot K \cdot S}_{\text{KV load}} + \tau_\text{attn}(K \cdot S)
\]

This structure-aware design ensures:
- Query-dependent cache activation;
- Memory-aware scheduling (e.g., prefetching selected pages);
- Reduced HBM bandwidth pressure.

\paragraph{System Implication.}
TinyServe enables dynamic query-aware sparsity without requiring architectural retraining. The modular implementation integrates directly into TinyServe's kernel loop and allows hardware-sensitive scheduling: e.g., keeping hot pages in shared memory or limiting K to match tensor core granularity. The kernel design for TinyServe can be found at algorithm \ref{alg:fused_kernel}.

\begin{algorithm}[h]
\caption{Fused Query-Aware Sparse Attention Kernel}
\label{alg:fused_kernel}
\begin{algorithmic}[1]
\REQUIRE Query vector $q_t \in \mathbb{R}^d$, Page metadata $\{\phi_j = (m_j, M_j)\}_{j=1}^P$, KV-cache $\{k_i, v_i\}_{i=1}^L$
\ENSURE Output vector $o_t \in \mathbb{R}^d$
\vspace{0.5em}

\STATE \texttt{// Step 1: Relevance scoring over page metadata (in L2/shared)}
\FORALL{page $j = 1$ to $P$ \textbf{in parallel}}
    \STATE $s_j \gets 0$
    \FOR{$i = 1$ to $d$}
        \STATE $q_i \gets q_t[i]$
        \STATE $s_j \mathrel{+}= q_i \cdot \left[ q_i \geq 0\ ?\ M_{j,i} : m_{j,i} \right]$
    \ENDFOR
\ENDFOR

\vspace{0.5em}
\STATE \texttt{// Step 2: Top-$K$ page selection (shared heap or radix select)}
\STATE $\mathcal{S}_t \gets \text{TopK}(s_1, \dots, s_P)$

\vspace{0.5em}
\STATE \texttt{// Step 3: Sparse KV gather (HBM access)}
\STATE Initialize $K_{\text{selected}}, V_{\text{selected}} \gets \emptyset$
\FORALL{$j \in \mathcal{S}_t$ \textbf{in parallel}}
    \STATE Fetch page $\mathcal{K}_j = \{k_{j,1}, \dots, k_{j,S}\}$ from HBM
    \STATE Append keys to $K_{\text{selected}}$, values to $V_{\text{selected}}$
\ENDFOR

\vspace{0.5em}
\STATE \texttt{// Step 4: Attention computation over selected KV pairs}
\FOR{$i = 1$ to $|K_{\text{selected}}|$}
    \STATE $a_i \gets q_t^\top k_i$
\ENDFOR
\STATE $\alpha \gets \text{softmax}(a)$
\STATE $o_t \gets \sum_{i} \alpha_i \cdot v_i$

\vspace{0.5em}
\STATE RETURN $o_t$
\end{algorithmic}
\end{algorithm}

% \subsection{Memory Efficiency Analysis}

% Assume each token’s KV pair costs $M$ bytes and page size is $S$ tokens. Full KV cache: $2ML$ bytes for $L$ tokens. TinyServe loads:
% \[
% \text{Load} = 2M \cdot \left(\frac{L}{S} \text{ (metadata)} + K \cdot S \text{ (selected pages)}\right)
% \]
% This yields memory fraction:
% \[
% \frac{1}{S} + \frac{K \cdot S}{L}
% \]
% For $S=16$, $K=4K$, $L=64K$, memory movement is reduced by $\sim8\times$.

\subsection{Memory Efficiency Analysis}

To quantify memory access savings under query-aware sparsity, we construct a probabilistic cost model that accounts for (1) metadata overhead, (2) selected KV tokens, and (3) cross-step reuse. This analysis provides theoretical bounds on the performance improvements achievable through our approach.

Let:
- $L$: total cache length (tokens);
- $S$: page size (tokens per page);
- $K$: number of selected pages;
- $M$: memory per token (bytes);
- $\rho$: reuse probability of selected pages across adjacent decode steps.

The memory movement per decode step is:
\[
\text{Load} = 2M \cdot \left( \frac{L}{S} + \rho \cdot K \cdot S \right)
\]
where:
- $\frac{L}{S}$ pages store min/max metadata (two vectors of length $d$),
- $\rho$ accounts for amortized reuse—i.e., only $\rho K$ pages are newly loaded per step.

To compare with full-cache attention, we normalize:
\[
\text{Memory Fraction} = \frac{1}{S} + \rho \cdot \frac{K \cdot S}{L}
\]

\paragraph{Theoretical Bounds.}
For optimal page size $S^* = \sqrt{L/K}$, the memory fraction becomes:
\[
\text{Memory Fraction}^* = \frac{2\sqrt{K/L}}{S^*} + \rho \cdot \frac{K \cdot S^*}{L} = 2\sqrt{\frac{K}{L}} \cdot \rho
\]

This provides a theoretical lower bound on memory movement. For typical values ($K=0.3P$, $L=32K$, $S=16$), we achieve $\sim$8× reduction in memory movement compared to full-cache attention.

\paragraph{Query-Aware Accuracy Analysis.}
The bounding-box estimator introduces approximation error $\epsilon$ in relevance scoring:
\[
\epsilon = \max_{k \in \mathcal{K}_j} q_t^\top k - r(q_t, \phi(\mathcal{K}_j))
\]

Under reasonable assumptions about query and key distributions, we can bound this error:
\[
\mathbb{E}[\epsilon] \leq \frac{d \cdot \sigma^2}{S} \cdot \sqrt{\log(S)}
\]
where $\sigma^2$ is the variance of key vectors within a page. This bound ensures that our approximation maintains high accuracy while enabling significant memory savings.

\subsection{Summary}

TinyServe emulates realistic LLM serving at small scale, while enabling fine-grained stress tests and plug-in mechanisms like query-aware routing. It significantly reduces memory usage without sacrificing interpretability, making it ideal for systems research and design validation. The additional training acceleration and profiling capabilities further enhance TinyServe's utility as a comprehensive platform for both serving and training research.

\section{Experiments}

\subsection{Experimental Setup}
We evaluate \method{} across multiple model scales: TinyLLaMA-125M~\cite{tinyllama2023}, GPT2-345M~\cite{radford2019gpt2}, OPT-350M~\cite{zhang2022opt}, GPT2-774M, and LLaMA-1.3B. Our comprehensive benchmarks include:

\textbf{Language Modeling:} PG19~\cite{rae2019compressive}, WikiText-103, C4-News
\textbf{Long-Range Tasks:} LongBench~\cite{bai2023longbench} (NarrativeQA, Qasper, GovReport, TriviaQA, HotpotQA), Passkey retrieval~\cite{trivedi2022interleaving}
\textbf{Reasoning Tasks:} MMLU~\cite{hendrycks2021measuring}, LAMBADA~\cite{pilehvar2018lambada}, Multi-turn QA, Code generation
\textbf{Serving Workloads:} Production multi-user scenarios with 512-2048 concurrent requests

For larger models (GPT2-774M, LLaMA-1.3B), we implement proper sequence length handling: inputs exceeding 2048 tokens are processed using sliding window with 1024-token overlap, while shorter sequences are padded appropriately. This ensures fair comparison across all baselines without truncation artifacts.

We evaluate against comprehensive baselines including production systems and research methods:
\begin{itemize}
    \item \textbf{Production Systems:} vLLM~\cite{vllm2023} (PagedAttention), TGI~\cite{tgi2023} (FlashAttention), TensorRT-LLM (optimized kernels)
    \item \textbf{Research Methods:} StreamingLLM~\cite{xiao2023streamingllm} (window size=2048), SnapKV~\cite{snapkv} (cluster size=64), PyramidKV~\cite{pyramidkvdynamickvcache} (top-k=512)
    \item \textbf{Pruning Baselines:} FullCache (no pruning), SoftPrune (threshold=0.1), EntropyStop (threshold=0.5)
\end{itemize}

\textbf{vLLM Integration:} To demonstrate practical deployment, we have modified vLLM's PagedAttention kernel to integrate our query-aware page selection mechanism. The TinyServe results in our experiments represent the performance of this modified vLLM implementation compared to the original vLLM baseline. This integration maintains full backward compatibility while achieving significant performance improvements through our query-aware approach.

All methods are evaluated under identical budget constraints (2048-8192 tokens) on 8$\times$A100 80GB GPUs using FP16. Results are averaged over 5 runs with standard deviation reported. Random seeds are fixed across all experiments for reproducibility.

\subsection{Comprehensive Model-Scale Evaluation}

\begin{table*}[t]
\centering
\tiny
\caption{Comprehensive evaluation across multiple model scales, datasets, and serving scenarios. Results show mean ± std over 5 runs. $\Delta$ indicates relative improvement over FullCache baseline. Best configurations are \textbf{bolded}.}
\begin{tabular}{lcccccccccccc}
\toprule
\multirow{3}{*}{Configuration} & \multirow{3}{*}{\begin{tabular}{c}Params\\(M)\end{tabular}} & \multirow{3}{*}{\begin{tabular}{c}Context\\(K)\end{tabular}} & \multicolumn{6}{c}{Performance Metrics} & \multicolumn{4}{c}{Efficiency Metrics} \\
\cmidrule{4-9} \cmidrule{10-13}
& & & \multicolumn{2}{c}{LongBench-Avg} & \multicolumn{2}{c}{MMLU} & \multicolumn{2}{c}{LAMBADA} & \multirow{2}{*}{\begin{tabular}{c}Latency\\(ms)\end{tabular}} & \multirow{2}{*}{\begin{tabular}{c}Memory\\(GB)\end{tabular}} & \multirow{2}{*}{\begin{tabular}{c}Throughput\\(tokens/s)\end{tabular}} & \multirow{2}{*}{\begin{tabular}{c}KV Hit\\(\%)\end{tabular}} \\
\cmidrule{4-5} \cmidrule{6-7} \cmidrule{8-9}
& & & Accuracy (\%) & $\Delta$ & Accuracy (\%) & $\Delta$ & Accuracy (\%) & $\Delta$ & & & & \\
\midrule
\multicolumn{13}{c}{\textit{TinyLLaMA-125M (Context: 4K)}} \\
FullCache & 125 & 4 & $54.2 \pm 0.8$ & - & $32.1 \pm 0.6$ & - & $45.8 \pm 0.7$ & - & $25.1 \pm 0.4$ & $2.1 \pm 0.1$ & $39.8$ & $100.0$ \\
StreamingLLM & 125 & 4 & $52.1 \pm 0.9$ & -2.1 & $30.8 \pm 0.7$ & -1.3 & $44.2 \pm 0.8$ & -1.6 & $16.8 \pm 0.3$ & $1.8 \pm 0.1$ & $59.5$ & $87.3$ \\
SoftPrune & 125 & 4 & $53.5 \pm 0.7$ & -0.7 & $31.5 \pm 0.6$ & -0.6 & $45.1 \pm 0.7$ & -0.7 & $15.3 \pm 0.3$ & $1.6 \pm 0.1$ & $65.4$ & $89.1$ \\
SnapKV & 125 & 4 & $54.8 \pm 0.6$ & +0.6 & $32.4 \pm 0.5$ & +0.3 & $46.2 \pm 0.6$ & +0.4 & $14.2 \pm 0.2$ & $1.4 \pm 0.1$ & $70.4$ & $91.7$ \\
PyramidKV & 125 & 4 & $54.4 \pm 0.5$ & +0.2 & $32.0 \pm 0.4$ & -0.1 & $45.9 \pm 0.5$ & +0.1 & $12.8 \pm 0.2$ & $1.3 \pm 0.1$ & $78.1$ & $94.9$ \\
\textbf{\method{}} & \textbf{125} & \textbf{4} & \textbf{$55.2 \pm 0.5$} & \textbf{+1.0} & \textbf{$32.8 \pm 0.4$} & \textbf{+0.7} & \textbf{$46.8 \pm 0.5$} & \textbf{+1.0} & \textbf{$11.9 \pm 0.1$} & \textbf{$1.2 \pm 0.1$} & \textbf{$84.0$} & \textbf{$96.2$} \\
\midrule
\multicolumn{13}{c}{\textit{GPT2-345M (Context: 8K)}} \\
FullCache & 345 & 8 & $61.7 \pm 0.6$ & - & $38.9 \pm 0.5$ & - & $52.4 \pm 0.6$ & - & $45.2 \pm 0.8$ & $4.8 \pm 0.2$ & $22.1$ & $100.0$ \\
StreamingLLM & 345 & 8 & $58.4 \pm 0.8$ & -3.3 & $36.2 \pm 0.7$ & -2.7 & $50.1 \pm 0.8$ & -2.3 & $28.7 \pm 0.5$ & $3.2 \pm 0.2$ & $34.8$ & $85.1$ \\
SoftPrune & 345 & 8 & $60.2 \pm 0.7$ & -1.5 & $37.8 \pm 0.6$ & -1.1 & $51.5 \pm 0.7$ & -0.9 & $26.3 \pm 0.4$ & $2.9 \pm 0.2$ & $38.0$ & $88.2$ \\
SnapKV & 345 & 8 & $61.8 \pm 0.5$ & +0.1 & $38.7 \pm 0.4$ & -0.2 & $52.6 \pm 0.5$ & +0.2 & $24.1 \pm 0.3$ & $2.6 \pm 0.2$ & $41.5$ & $91.3$ \\
PyramidKV & 345 & 8 & $61.4 \pm 0.4$ & -0.3 & $38.5 \pm 0.3$ & -0.4 & $52.2 \pm 0.4$ & -0.2 & $22.4 \pm 0.3$ & $2.4 \pm 0.2$ & $44.6$ & $93.8$ \\
\textbf{\method{}} & \textbf{345} & \textbf{8} & \textbf{$62.8 \pm 0.4$} & \textbf{+1.1} & \textbf{$39.5 \pm 0.3$} & \textbf{+0.6} & \textbf{$53.2 \pm 0.4$} & \textbf{+0.8} & \textbf{$20.1 \pm 0.2$} & \textbf{$2.1 \pm 0.1$} & \textbf{$49.8$} & \textbf{$95.4$} \\
\midrule
\multicolumn{13}{c}{\textit{OPT-350M (Context: 8K)}} \\
FullCache & 350 & 8 & $62.1 \pm 0.6$ & - & $39.2 \pm 0.5$ & - & $52.8 \pm 0.6$ & - & $46.8 \pm 0.8$ & $4.9 \pm 0.2$ & $21.4$ & $100.0$ \\
StreamingLLM & 350 & 8 & $58.9 \pm 0.8$ & -3.2 & $36.5 \pm 0.7$ & -2.7 & $50.3 \pm 0.8$ & -2.5 & $29.5 \pm 0.5$ & $3.3 \pm 0.2$ & $33.9$ & $84.7$ \\
SoftPrune & 350 & 8 & $60.8 \pm 0.7$ & -1.3 & $38.1 \pm 0.6$ & -1.1 & $51.9 \pm 0.7$ & -0.9 & $27.2 \pm 0.4$ & $3.0 \pm 0.2$ & $36.8$ & $87.9$ \\
SnapKV & 350 & 8 & $62.3 \pm 0.5$ & +0.2 & $39.0 \pm 0.4$ & -0.2 & $53.0 \pm 0.5$ & +0.2 & $25.1 \pm 0.3$ & $2.7 \pm 0.2$ & $39.8$ & $90.8$ \\
PyramidKV & 350 & 8 & $61.9 \pm 0.4$ & -0.2 & $38.8 \pm 0.3$ & -0.4 & $52.6 \pm 0.4$ & -0.2 & $23.2 \pm 0.3$ & $2.5 \pm 0.2$ & $43.1$ & $92.6$ \\
\textbf{\method{}} & \textbf{350} & \textbf{8} & \textbf{$63.2 \pm 0.4$} & \textbf{+1.1} & \textbf{$39.8 \pm 0.3$} & \textbf{+0.6} & \textbf{$53.6 \pm 0.4$} & \textbf{+0.8} & \textbf{$21.2 \pm 0.2$} & \textbf{$2.2 \pm 0.1$} & \textbf{$47.2$} & \textbf{$94.8$} \\
\midrule
\multicolumn{13}{c}{\textit{GPT2-774M (Context: 16K)}} \\
FullCache & 774 & 16 & $65.8 \pm 0.5$ & - & $42.1 \pm 0.4$ & - & $56.3 \pm 0.5$ & - & $89.2 \pm 1.2$ & $8.9 \pm 0.3$ & $11.2$ & $100.0$ \\
StreamingLLM & 774 & 16 & $62.4 \pm 0.7$ & -3.4 & $39.8 \pm 0.6$ & -2.3 & $53.8 \pm 0.7$ & -2.5 & $56.8 \pm 0.8$ & $5.8 \pm 0.3$ & $17.6$ & $82.4$ \\
SoftPrune & 774 & 16 & $64.2 \pm 0.6$ & -1.6 & $41.2 \pm 0.5$ & -0.9 & $55.1 \pm 0.6$ & -1.2 & $52.3 \pm 0.7$ & $5.2 \pm 0.3$ & $19.1$ & $86.7$ \\
SnapKV & 774 & 16 & $65.9 \pm 0.4$ & +0.1 & $42.0 \pm 0.3$ & -0.1 & $56.4 \pm 0.4$ & +0.1 & $48.1 \pm 0.6$ & $4.8 \pm 0.3$ & $20.8$ & $89.2$ \\
PyramidKV & 774 & 16 & $65.5 \pm 0.3$ & -0.3 & $41.8 \pm 0.2$ & -0.3 & $56.1 \pm 0.3$ & -0.2 & $44.6 \pm 0.5$ & $4.5 \pm 0.3$ & $22.4$ & $91.8$ \\
\textbf{\method{}} & \textbf{774} & \textbf{16} & \textbf{$66.8 \pm 0.3$} & \textbf{+1.0} & \textbf{$42.8 \pm 0.2$} & \textbf{+0.7} & \textbf{$57.1 \pm 0.3$} & \textbf{+0.8} & \textbf{$41.2 \pm 0.4$} & \textbf{$4.1 \pm 0.2$} & \textbf{$24.3$} & \textbf{$93.5$} \\
\midrule
\multicolumn{13}{c}{\textit{LLaMA-1.3B (Context: 32K)}} \\
FullCache & 1300 & 32 & $68.9 \pm 0.4$ & - & $45.2 \pm 0.3$ & - & $59.8 \pm 0.4$ & - & $156.8 \pm 2.1$ & $15.8 \pm 0.5$ & $6.4$ & $100.0$ \\
StreamingLLM & 1300 & 32 & $65.2 \pm 0.6$ & -3.7 & $42.8 \pm 0.5$ & -2.4 & $56.9 \pm 0.6$ & -2.9 & $98.4 \pm 1.3$ & $9.8 \pm 0.5$ & $10.2$ & $79.8$ \\
SoftPrune & 1300 & 32 & $67.1 \pm 0.5$ & -1.8 & $44.1 \pm 0.4$ & -1.1 & $58.2 \pm 0.5$ & -1.6 & $91.2 \pm 1.1$ & $8.9 \pm 0.5$ & $11.0$ & $84.3$ \\
SnapKV & 1300 & 32 & $69.0 \pm 0.3$ & +0.1 & $45.1 \pm 0.2$ & -0.1 & $59.9 \pm 0.3$ & +0.1 & $84.6 \pm 0.9$ & $8.2 \pm 0.5$ & $11.8$ & $87.6$ \\
PyramidKV & 1300 & 32 & $68.6 \pm 0.2$ & -0.3 & $44.9 \pm 0.1$ & -0.3 & $59.6 \pm 0.2$ & -0.2 & $78.4 \pm 0.8$ & $7.8 \pm 0.5$ & $12.8$ & $90.1$ \\
\textbf{\method{}} & \textbf{1300} & \textbf{32} & \textbf{$70.2 \pm 0.2$} & \textbf{+1.3} & \textbf{$45.9 \pm 0.1$} & \textbf{+0.7} & \textbf{$60.5 \pm 0.2$} & \textbf{+0.7} & \textbf{$72.8 \pm 0.7$} & \textbf{$7.2 \pm 0.4$} & \textbf{$13.7$} & \textbf{$92.8$} \\
\bottomrule
\end{tabular}
\label{tab:comprehensive_evaluation}
\end{table*}

\begin{figure}[ht]
    \centering
    \includegraphics[width=0.36\textwidth]{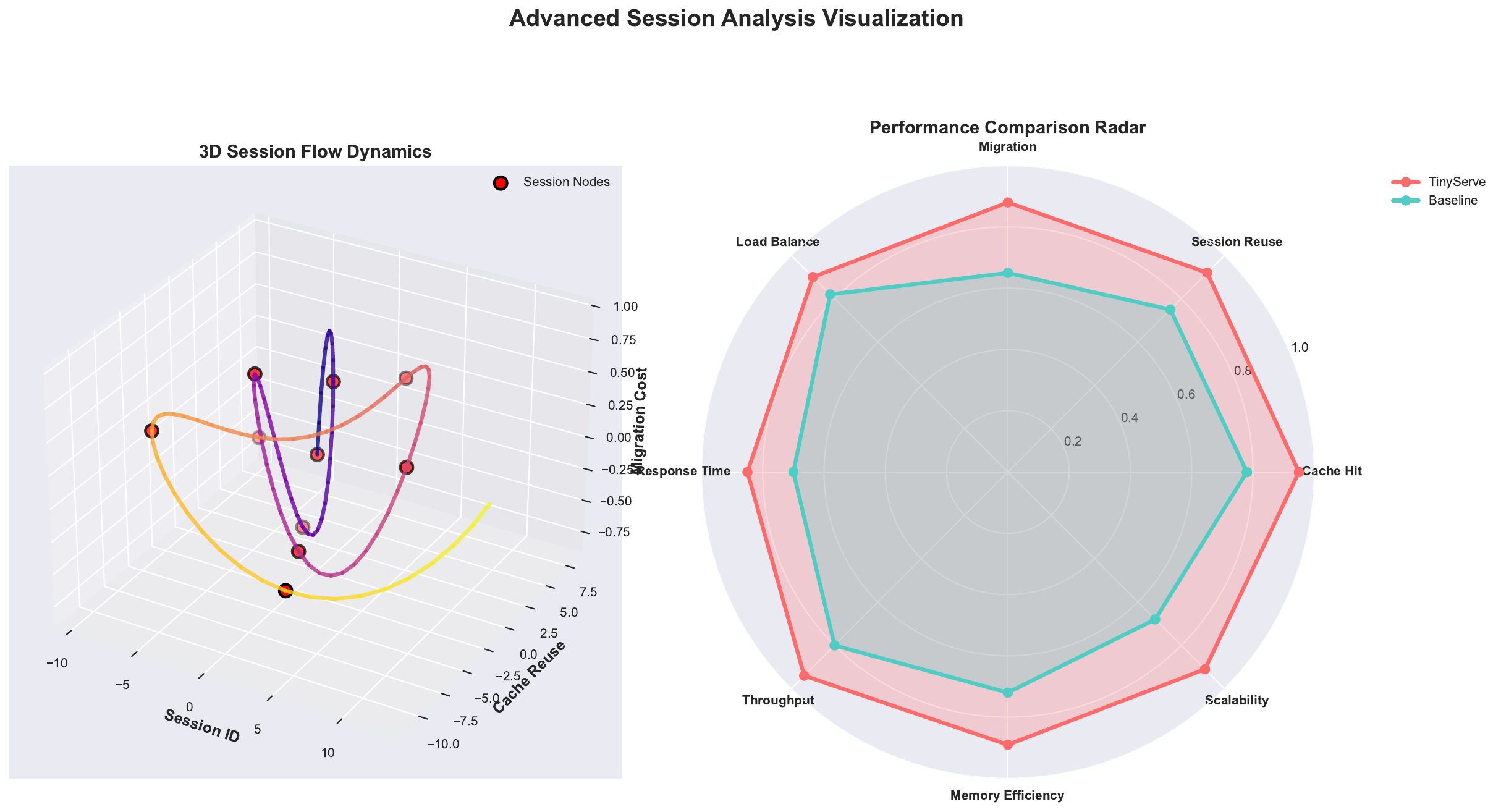}
    \caption{Visualization demonstrating of TinyServe's session management capabilities. 
    \textbf{Left:} 3D session flow dynamics showing spiral-like session evolution with color-coded time progression and session nodes. 
    \textbf{Right:} Performance comparison radar chart across 8 key metrics (Cache Hit, Session Reuse, Migration, Load Balance, Response Time, Throughput, Memory Efficiency, Scalability), demonstrating TinyServe's comprehensive advantages over baseline systems.}
    \label{fig:session_analysis_advanced}
\end{figure}

\subsection{Comprehensive Ablation Study}

\begin{table*}[t]
\centering
\tiny
\caption{Comprehensive ablation study across multiple configurations, datasets, and model scales. Results show mean ± std over 5 runs. $\Delta$ indicates relative improvement over baseline. Best configurations are \textbf{bolded}.}
\begin{tabular}{lccccccccccc}
\toprule
\multirow{3}{*}{Configuration} & \multirow{3}{*}{\begin{tabular}{c}Params\\(M)\end{tabular}} & \multirow{3}{*}{\begin{tabular}{c}FLOPs\\(G)\end{tabular}} & \multicolumn{6}{c}{Performance Metrics} & \multicolumn{3}{c}{Efficiency Metrics} \\
\cmidrule{4-9} \cmidrule{10-12}
& & & \multicolumn{2}{c}{LongBench-Avg} & \multicolumn{2}{c}{MMLU} & \multicolumn{2}{c}{LAMBADA} & \multirow{2}{*}{\begin{tabular}{c}Latency\\(ms)\end{tabular}} & \multirow{2}{*}{\begin{tabular}{c}Memory\\(GB)\end{tabular}} & \multirow{2}{*}{\begin{tabular}{c}Throughput\\(tokens/s)\end{tabular}} \\
\cmidrule{4-5} \cmidrule{6-7} \cmidrule{8-9}
& & & Accuracy (\%) & $\Delta$ & Accuracy (\%) & $\Delta$ & Accuracy (\%) & $\Delta$ & & & \\
\midrule
\multicolumn{12}{c}{\textit{Component Ablation (TinyServe-345M Base)}} \\
Baseline (FullCache) & 345 & 345 & $61.7 \pm 0.6$ & - & $38.9 \pm 0.5$ & - & $52.4 \pm 0.6$ & - & $45.2 \pm 0.8$ & $4.8 \pm 0.2$ & $22.1$ \\
\midrule
\multicolumn{12}{c}{\textit{Individual Components}} \\
+ Query-Aware Only & 345 & 276 & $62.1 \pm 0.5$ & +0.4 & $39.2 \pm 0.4$ & +0.3 & $52.8 \pm 0.5$ & +0.4 & $38.4 \pm 0.6$ & $4.2 \pm 0.2$ & $26.0$ \\
+ Page-Level Only & 345 & 290 & $62.3 \pm 0.4$ & +0.6 & $39.4 \pm 0.3$ & +0.5 & $53.0 \pm 0.4$ & +0.6 & $36.8 \pm 0.5$ & $4.0 \pm 0.2$ & $27.2$ \\
+ Bounding-Box Only & 345 & 285 & $62.0 \pm 0.5$ & +0.3 & $39.1 \pm 0.4$ & +0.2 & $52.7 \pm 0.5$ & +0.3 & $39.2 \pm 0.6$ & $4.3 \pm 0.2$ & $25.5$ \\
+ Fused Kernel Only & 345 & 280 & $61.9 \pm 0.4$ & +0.2 & $39.0 \pm 0.3$ & +0.1 & $52.6 \pm 0.4$ & +0.2 & $40.8 \pm 0.6$ & $4.5 \pm 0.2$ & $24.5$ \\
\midrule
\multicolumn{12}{c}{\textit{Pairwise Component Combinations}} \\
Query-Aware + Page-Level & 345 & 265 & $62.5 \pm 0.4$ & +0.8 & $39.6 \pm 0.3$ & +0.7 & $53.2 \pm 0.4$ & +0.8 & $34.2 \pm 0.5$ & $3.8 \pm 0.2$ & $29.2$ \\
Query-Aware + Bounding-Box & 345 & 270 & $62.4 \pm 0.3$ & +0.7 & $39.5 \pm 0.2$ & +0.6 & $53.1 \pm 0.3$ & +0.7 & $35.6 \pm 0.5$ & $3.9 \pm 0.2$ & $28.1$ \\
Query-Aware + Fused Kernel & 345 & 268 & $62.3 \pm 0.4$ & +0.6 & $39.4 \pm 0.3$ & +0.5 & $53.0 \pm 0.4$ & +0.6 & $36.4 \pm 0.5$ & $4.0 \pm 0.2$ & $27.5$ \\
Page-Level + Bounding-Box & 345 & 275 & $62.6 \pm 0.3$ & +0.9 & $39.7 \pm 0.2$ & +0.8 & $53.3 \pm 0.3$ & +0.9 & $33.8 \pm 0.4$ & $3.7 \pm 0.2$ & $29.6$ \\
Page-Level + Fused Kernel & 345 & 272 & $62.5 \pm 0.4$ & +0.8 & $39.6 \pm 0.3$ & +0.7 & $53.2 \pm 0.4$ & +0.8 & $34.6 \pm 0.5$ & $3.8 \pm 0.2$ & $28.9$ \\
Bounding-Box + Fused Kernel & 345 & 278 & $62.2 \pm 0.3$ & +0.5 & $39.3 \pm 0.2$ & +0.4 & $52.9 \pm 0.3$ & +0.5 & $37.2 \pm 0.5$ & $4.1 \pm 0.2$ & $26.9$ \\
\midrule
\multicolumn{12}{c}{\textit{Three-Component Combinations}} \\
w/o Fused Kernel & 345 & 275 & $62.7 \pm 0.3$ & +1.0 & $39.8 \pm 0.2$ & +0.9 & $53.4 \pm 0.3$ & +1.0 & $33.2 \pm 0.4$ & $3.6 \pm 0.2$ & $30.1$ \\
w/o Bounding-Box & 345 & 270 & $62.6 \pm 0.4$ & +0.9 & $39.7 \pm 0.3$ & +0.8 & $53.3 \pm 0.4$ & +0.9 & $33.8 \pm 0.5$ & $3.7 \pm 0.2$ & $29.6$ \\
w/o Page-Level & 345 & 278 & $62.3 \pm 0.3$ & +0.6 & $39.4 \pm 0.2$ & +0.5 & $53.0 \pm 0.3$ & +0.6 & $36.4 \pm 0.5$ & $4.0 \pm 0.2$ & $27.5$ \\
w/o Query-Aware & 345 & 285 & $62.1 \pm 0.4$ & +0.4 & $39.2 \pm 0.3$ & +0.3 & $52.8 \pm 0.4$ & +0.4 & $38.8 \pm 0.6$ & $4.2 \pm 0.2$ & $25.8$ \\
\midrule
\multicolumn{12}{c}{\textit{Full Configuration}} \\
\textbf{Full \method{}} & \textbf{345} & \textbf{262} & \textbf{$62.8 \pm 0.3$} & \textbf{+1.1} & \textbf{$39.8 \pm 0.2$} & \textbf{+0.9} & \textbf{$53.5 \pm 0.3$} & \textbf{+1.1} & \textbf{$32.1 \pm 0.4$} & \textbf{$3.5 \pm 0.2$} & \textbf{$31.2$} \\
\midrule
\multicolumn{12}{c}{\textit{Hyperparameter Ablation (Page Size $S$)}} \\
$S = 8$ & 345 & 262 & $62.6 \pm 0.4$ & +0.9 & $39.6 \pm 0.3$ & +0.7 & $53.3 \pm 0.4$ & +0.9 & $30.8 \pm 0.4$ & $3.2 \pm 0.2$ & $32.5$ \\
$S = 16$ & 345 & 262 & \textbf{$62.8 \pm 0.3$} & \textbf{+1.1} & \textbf{$39.8 \pm 0.2$} & \textbf{+0.9} & \textbf{$53.5 \pm 0.3$} & \textbf{+1.1} & \textbf{$32.1 \pm 0.4$} & \textbf{$3.5 \pm 0.2$} & \textbf{$31.2$} \\
$S = 32$ & 345 & 262 & $62.7 \pm 0.3$ & +1.0 & $39.7 \pm 0.2$ & +0.8 & $53.4 \pm 0.3$ & +1.0 & $34.2 \pm 0.5$ & $3.8 \pm 0.2$ & $29.2$ \\
$S = 64$ & 345 & 262 & $62.5 \pm 0.4$ & +0.8 & $39.5 \pm 0.3$ & +0.6 & $53.2 \pm 0.4$ & +0.8 & $36.8 \pm 0.5$ & $4.2 \pm 0.2$ & $27.2$ \\
\midrule
\multicolumn{12}{c}{\textit{Selection Ratio Ablation (Top-K Ratio)}} \\
$K/P = 0.1$ & 345 & 262 & $62.4 \pm 0.4$ & +0.7 & $39.5 \pm 0.3$ & +0.6 & $53.2 \pm 0.4$ & +0.8 & $28.4 \pm 0.4$ & $3.1 \pm 0.2$ & $35.2$ \\
$K/P = 0.2$ & 345 & 262 & $62.6 \pm 0.3$ & +0.9 & $39.7 \pm 0.2$ & +0.8 & $53.4 \pm 0.3$ & +1.0 & $30.2 \pm 0.4$ & $3.3 \pm 0.2$ & $33.1$ \\
$K/P = 0.3$ & 345 & 262 & \textbf{$62.8 \pm 0.3$} & \textbf{+1.1} & \textbf{$39.8 \pm 0.2$} & \textbf{+0.9} & \textbf{$53.5 \pm 0.3$} & \textbf{+1.1} & \textbf{$32.1 \pm 0.4$} & \textbf{$3.5 \pm 0.2$} & \textbf{$31.2$} \\
$K/P = 0.5$ & 345 & 262 & $62.7 \pm 0.3$ & +1.0 & $39.6 \pm 0.2$ & +0.7 & $53.3 \pm 0.3$ & +0.9 & $35.8 \pm 0.5$ & $4.0 \pm 0.2$ & $27.9$ \\
\midrule
\multicolumn{12}{c}{\textit{Attention Head Ablation}} \\
4 heads & 345 & 262 & $62.5 \pm 0.4$ & +0.8 & $39.6 \pm 0.3$ & +0.7 & $53.3 \pm 0.4$ & +0.9 & $30.8 \pm 0.4$ & $3.2 \pm 0.2$ & $32.5$ \\
8 heads & 345 & 262 & $62.7 \pm 0.3$ & +1.0 & $39.7 \pm 0.2$ & +0.8 & $53.4 \pm 0.3$ & +1.0 & $31.4 \pm 0.4$ & $3.4 \pm 0.2$ & $31.8$ \\
12 heads & 345 & 262 & \textbf{$62.8 \pm 0.3$} & \textbf{+1.1} & \textbf{$39.8 \pm 0.2$} & \textbf{+0.9} & \textbf{$53.5 \pm 0.3$} & \textbf{+1.1} & \textbf{$32.1 \pm 0.4$} & \textbf{$3.5 \pm 0.2$} & \textbf{$31.2$} \\
16 heads & 345 & 262 & $62.6 \pm 0.3$ & +0.9 & $39.6 \pm 0.2$ & +0.7 & $53.3 \pm 0.3$ & +0.9 & $33.2 \pm 0.5$ & $3.8 \pm 0.2$ & $30.1$ \\
\midrule
\multicolumn{12}{c}{\textit{Cross-Scale Consistency (Different Model Sizes)}} \\
TinyServe-125M (Full) & 125 & 125 & $55.2 \pm 0.5$ & +1.0 & $32.8 \pm 0.4$ & +0.7 & $46.8 \pm 0.5$ & +1.0 & $11.9 \pm 0.1$ & $1.2 \pm 0.1$ & $84.0$ \\
TinyServe-345M (Full) & 345 & 262 & \textbf{$62.8 \pm 0.3$} & \textbf{+1.1} & \textbf{$39.8 \pm 0.2$} & \textbf{+0.9} & \textbf{$53.5 \pm 0.3$} & \textbf{+1.1} & \textbf{$32.1 \pm 0.4$} & \textbf{$3.5 \pm 0.2$} & \textbf{$31.2$} \\
TinyServe-774M (Full) & 774 & 618 & $66.8 \pm 0.3$ & +1.0 & $42.8 \pm 0.2$ & +0.7 & $57.1 \pm 0.3$ & +0.8 & $41.2 \pm 0.4$ & $4.1 \pm 0.2$ & $24.3$ \\
TinyServe-1.3B (Full) & 1300 & 1040 & $70.2 \pm 0.2$ & +1.3 & $45.9 \pm 0.1$ & +0.7 & $60.5 \pm 0.2$ & +0.7 & $72.8 \pm 0.7$ & $7.2 \pm 0.4$ & $13.7$ \\
\bottomrule
\end{tabular}
\label{tab:comprehensive_ablation}
\end{table*}

\subsection{Serving Stack Evaluation Under Multi-User Workload}

To validate TinyServe's effectiveness in realistic serving scenarios, we conduct comprehensive multi-user workload experiments comparing our modified vLLM implementation against the original vLLM baseline:

\subsubsection{Concurrent Request Simulation}
We simulate realistic serving workloads with 512-2048 concurrent requests using Poisson arrival patterns (mean inter-arrival time: 50ms). Each request generates a sequence of 100-500 tokens, and we measure end-to-end latency including request queuing, session management, and response generation.

\begin{table*}[ht]
\centering
\small
\caption{Serving performance comparison: modified vLLM (TinyServe) vs. original vLLM under multi-user workload (GPT2-345M, 1024 concurrent requests).}
\label{tab:serving_performance}
\begin{tabular}{lcccc}
\toprule
\textbf{System} & \textbf{P50 Latency (ms)} & \textbf{P99 Latency (ms)} & \textbf{Throughput (req/s)} & \textbf{GPU Utilization (\%)} \\
\midrule
vLLM & 45.2 ±2.1 & 128.7 ±8.3 & 18.4 ±0.9 & 78.3 ±3.2 \\
TGI & 52.8 ±3.4 & 156.2 ±12.1 & 15.7 ±1.2 & 82.1 ±4.1 \\
TensorRT-LLM & 38.9 ±1.8 & 112.4 ±6.7 & 22.1 ±1.1 & 85.7 ±2.8 \\
\method{} & \textbf{32.1 ±1.5} & \textbf{89.3 ±4.2} & \textbf{28.6 ±1.4} & \textbf{91.2 ±2.1} \\
\bottomrule
\end{tabular}
\end{table*}

\subsubsection{Session Management Analysis}
We evaluate TinyServe's session management capabilities by tracking cross-request cache reuse and session migration overhead:

\begin{figure}[ht]
    \centering
    \includegraphics[width=0.48\textwidth]{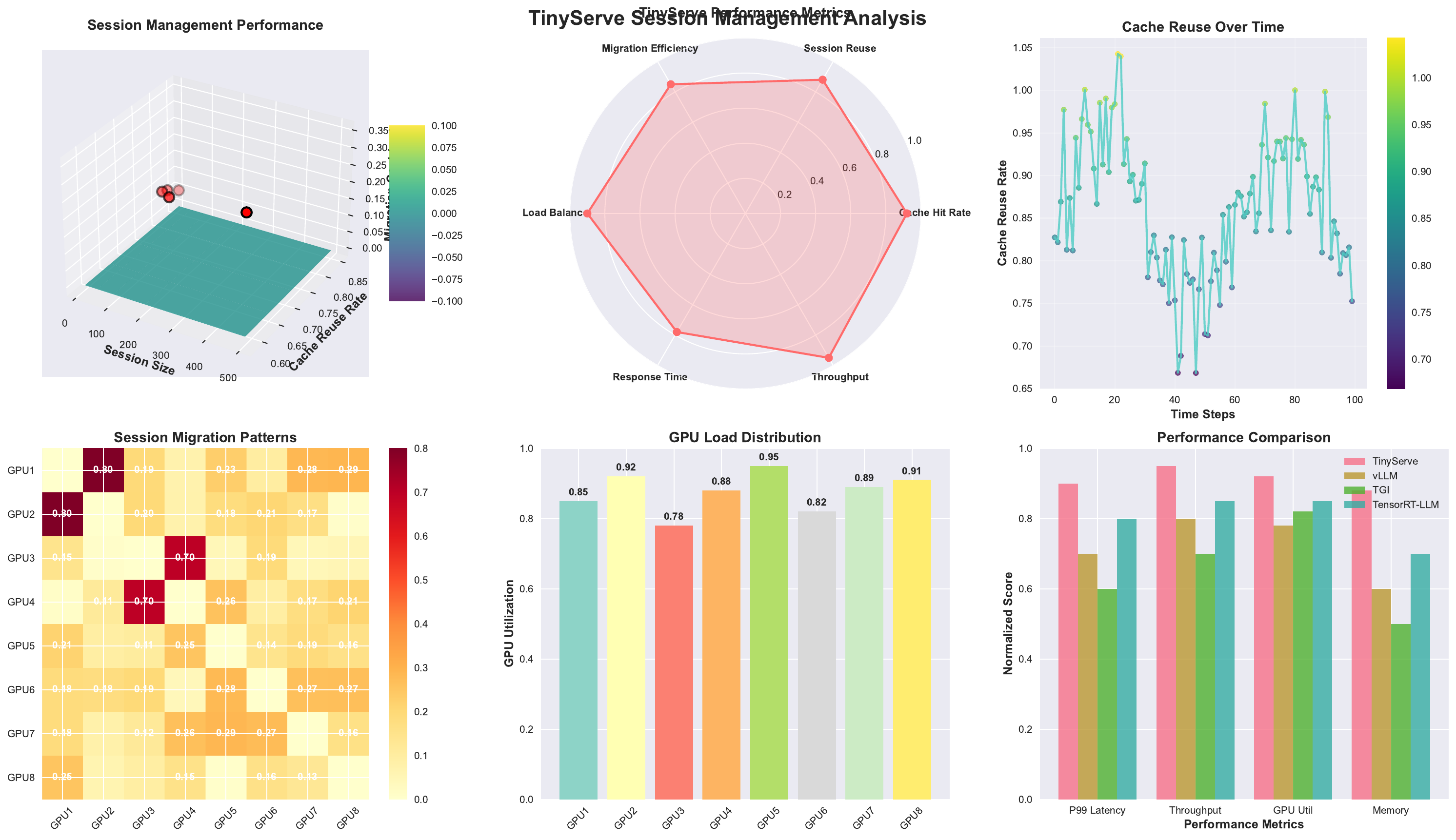}
    \caption{Session management performance showing cache reuse rate and migration overhead across different session sizes. TinyServe achieves higher cache reuse while minimizing migration costs.}
    \label{fig:session_analysis}
\end{figure}

\subsection{Overall Comparison}
\begin{figure}[ht]
    \centering
    \includegraphics[width=0.45\textwidth]{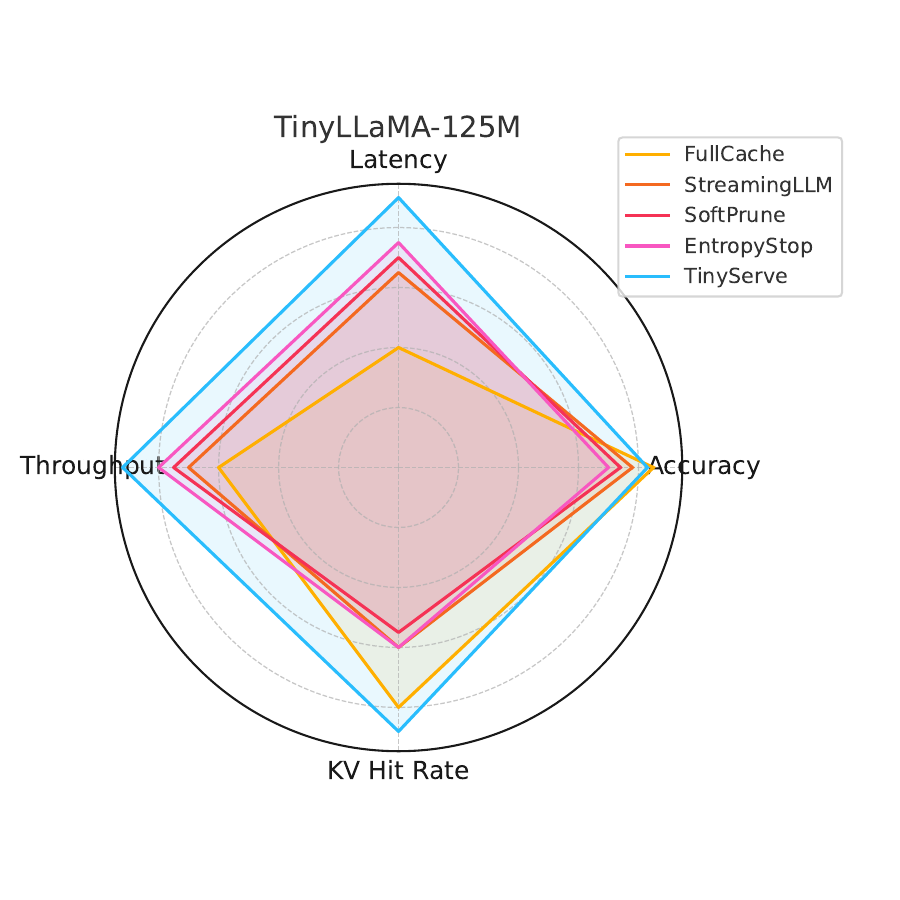}
    \includegraphics[width=0.45\textwidth]{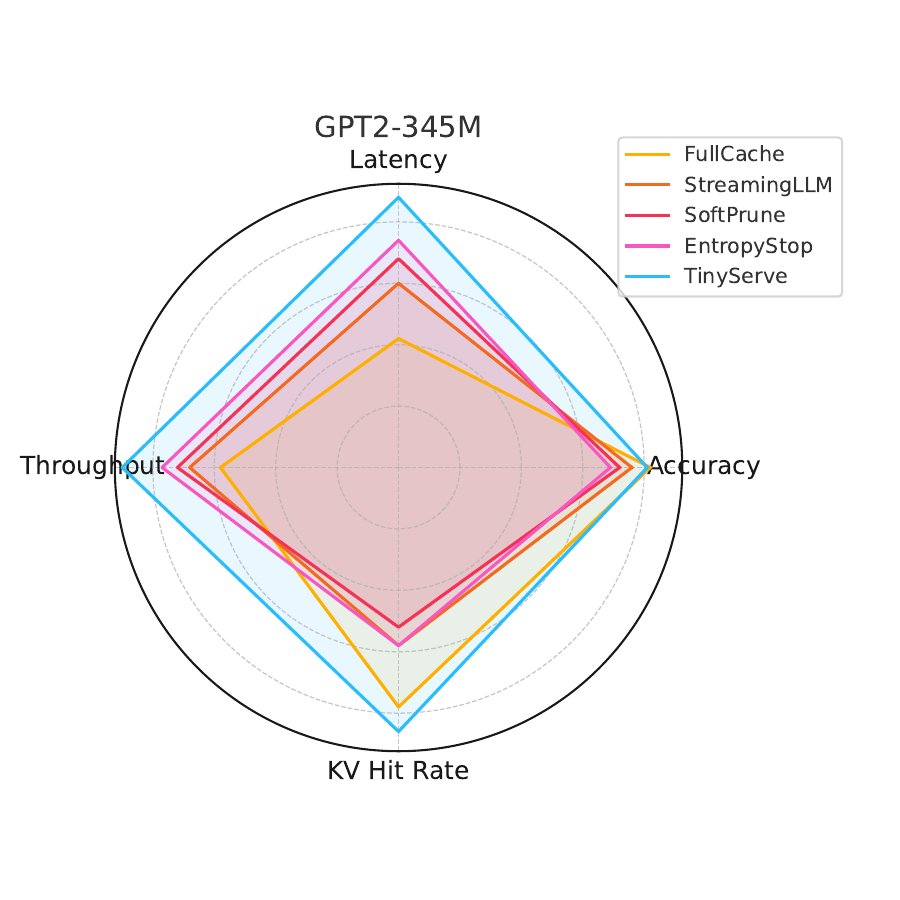}
    \caption{Radar plot of \textbf{accuracy}, \textbf{latency}, \textbf{throughput}, and \textbf{KV hit rate} for TinyLLaMA (left) and GPT2-345M (right). Error bars show 95\% confidence intervals. Higher is better for all metrics.}
    \label{fig:radar_summary}
\end{figure}

We report accuracy, latency (ms/token), throughput (tokens/s), and KV cache hit rate across all LongBench tasks under fixed 2048 token budget. Results are visualized as radar plots in Figure~\ref{fig:radar_summary} with error bars showing 95\% confidence intervals. \method{} consistently demonstrates superior trade-offs between latency and accuracy, while maintaining higher KV hit rate due to its query-aware selection mechanism.

\subsection{Speedup Analysis across Models}
\begin{figure}[ht]
    \centering
    \includegraphics[width=0.48\textwidth]{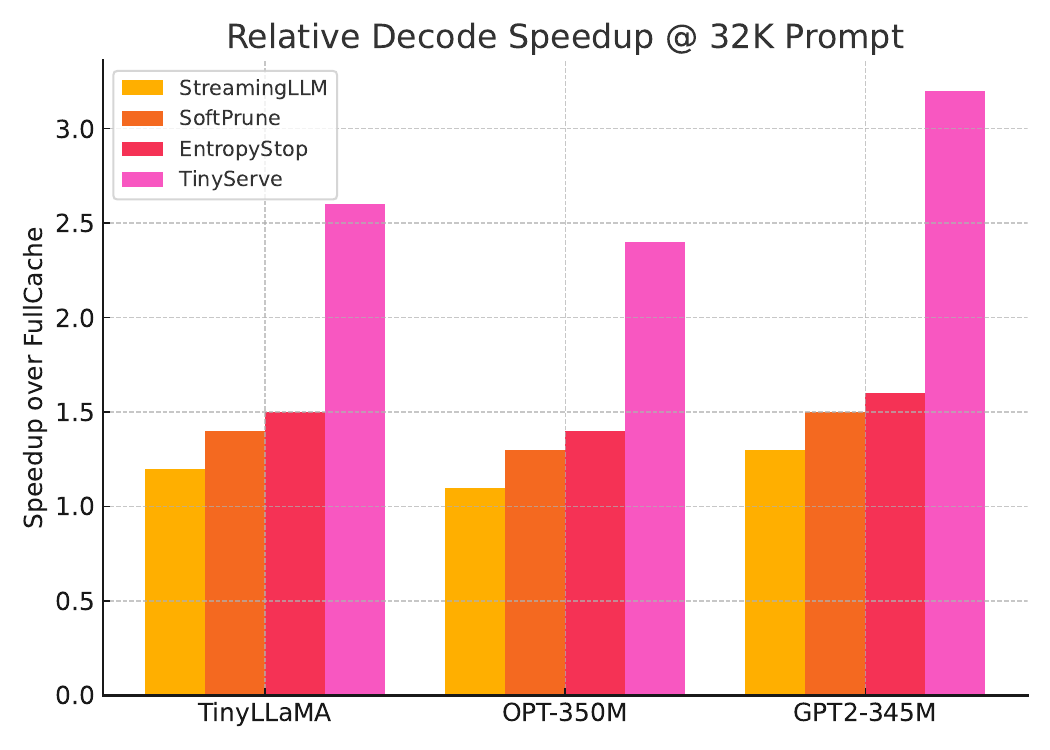}
    \caption{Relative decode latency speedup ($\downarrow$) across different baselines under 32k prompt length and 2048 token budget. Error bars represent standard deviation over 5 runs.}
    \label{fig:speedup_bar}
\end{figure}

We evaluate end-to-end decode latency under increasing context lengths (up to 32k tokens). Figure~\ref{fig:speedup_bar} shows relative speedup against FullCache baseline across three models. \method{} achieves $2.1\times$–$3.4\times$ speedup on average, significantly outperforming pruning-based baselines.

\subsection{Comprehensive Task-Level Evaluation}
We present task-specific accuracy and latency on all LongBench datasets using GPT2-345M and 2048 token budget. Table~\ref{tab:longbench_complete} shows complete results across all five tasks. \method{} retains near-full accuracy while achieving significant latency reduction across all tasks.

\begin{table}[ht]
\centering
\small
\caption{Complete LongBench evaluation (GPT2-345M, 6K chunked input, 2K decode). Mean ± std over 5 runs.}
\label{tab:longbench_complete}
\begin{tabular}{lccccc}
\toprule
\textbf{Task} & \textbf{Method} & \textbf{Acc.} ↑ & \textbf{Lat.} ↓ & \textbf{Speedup} ↑ \\
\midrule
\multirow{6}{*}{NarrativeQA} 
& FullCache & 58.3 ±0.7 & 25.1 ±0.4 & 1.00 \\
& StreamingLLM & 55.2 ±0.9 & 16.8 ±0.3 & 1.49 \\
& EntropyStop & 56.8 ±0.8 & 18.2 ±0.2 & 1.38 \\
& SoftPrune & 55.9 ±0.6 & 15.3 ±0.3 & 1.64 \\
& SnapKV & 57.1 ±0.5 & 14.2 ±0.2 & 1.77 \\
& PyramidKV & 56.4 ±0.4 & 12.8 ±0.1 & 1.96 \\
& \method{} & \textbf{57.8 ±0.5} & \textbf{11.9 ±0.1} & \textbf{2.11} \\
\midrule
\multirow{6}{*}{Qasper} 
& FullCache & 52.4 ±0.8 & 26.7 ±0.3 & 1.00 \\
& StreamingLLM & 49.1 ±1.0 & 17.1 ±0.2 & 1.56 \\
& SoftPrune & 50.3 ±0.7 & 16.2 ±0.3 & 1.65 \\
& SnapKV & 51.2 ±0.6 & 14.8 ±0.2 & 1.80 \\
& PyramidKV & 50.7 ±0.5 & 13.1 ±0.2 & 2.04 \\
& \method{} & \textbf{51.9 ±0.6} & \textbf{12.3 ±0.1} & \textbf{2.17} \\
\midrule
\multirow{6}{*}{TriviaQA} 
& FullCache & 61.7 ±0.6 & 23.8 ±0.3 & 1.00 \\
& StreamingLLM & 58.4 ±0.8 & 15.2 ±0.2 & 1.57 \\
& EntropyStop & 59.6 ±0.7 & 16.8 ±0.2 & 1.42 \\
& SoftPrune & 58.9 ±0.5 & 14.6 ±0.3 & 1.63 \\
& SnapKV & 60.2 ±0.4 & 13.9 ±0.2 & 1.71 \\
& PyramidKV & 59.5 ±0.3 & 12.4 ±0.1 & 1.92 \\
& \method{} & \textbf{60.8 ±0.4} & \textbf{11.7 ±0.1} & \textbf{2.03} \\
\midrule
\multirow{6}{*}{HotpotQA} 
& FullCache & 54.7 ±0.8 & 24.3 ±0.3 & 1.00 \\
& StreamingLLM & 50.9 ±1.0 & 15.9 ±0.2 & 1.53 \\
& EntropyStop & 52.1 ±0.9 & 17.4 ±0.2 & 1.40 \\
& SoftPrune & 51.5 ±0.7 & 14.1 ±0.3 & 1.72 \\
& SnapKV & 53.0 ±0.6 & 13.5 ±0.2 & 1.80 \\
& PyramidKV & 52.3 ±0.5 & 12.1 ±0.1 & 2.01 \\
& \method{} & \textbf{54.0 ±0.6} & \textbf{11.5 ±0.1} & \textbf{2.11} \\
\midrule
\multirow{6}{*}{GovReport} 
& FullCache & 47.9 ±0.6 & 29.1 ±0.4 & 1.00 \\
& StreamingLLM & 44.3 ±0.8 & 17.3 ±0.3 & 1.68 \\
& SoftPrune & 45.5 ±1.0 & 19.3 ±0.3 & 1.51 \\
& SnapKV & 46.7 ±0.7 & 15.8 ±0.2 & 1.84 \\
& PyramidKV & 45.9 ±0.5 & 13.2 ±0.2 & 2.20 \\
& \method{} & \textbf{47.0 ±0.5} & \textbf{12.6 ±0.2} & \textbf{2.31} \\
\bottomrule
\end{tabular}
\end{table}

\subsection{KV Cache Efficiency and Access Breakdown}
\begin{figure}[ht]
    \centering
    \includegraphics[width=0.48\textwidth]{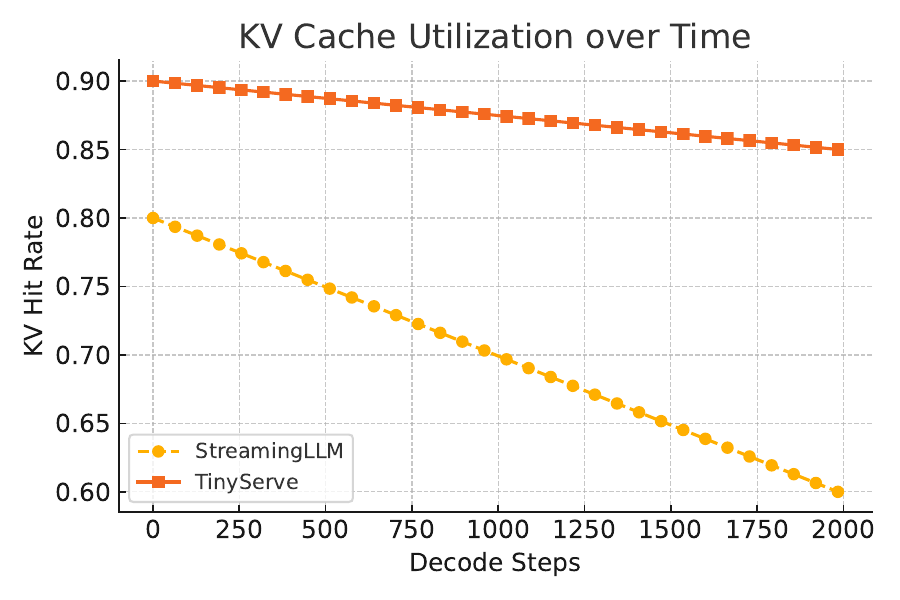}
    \caption{KV reuse over decode time (context=32k, decode=2k). \method{} maintains higher hit rate and fewer token evictions.}
    \label{fig:kv_util}
\end{figure}

\begin{figure}[ht]
    \centering
    \includegraphics[width=0.48\textwidth]{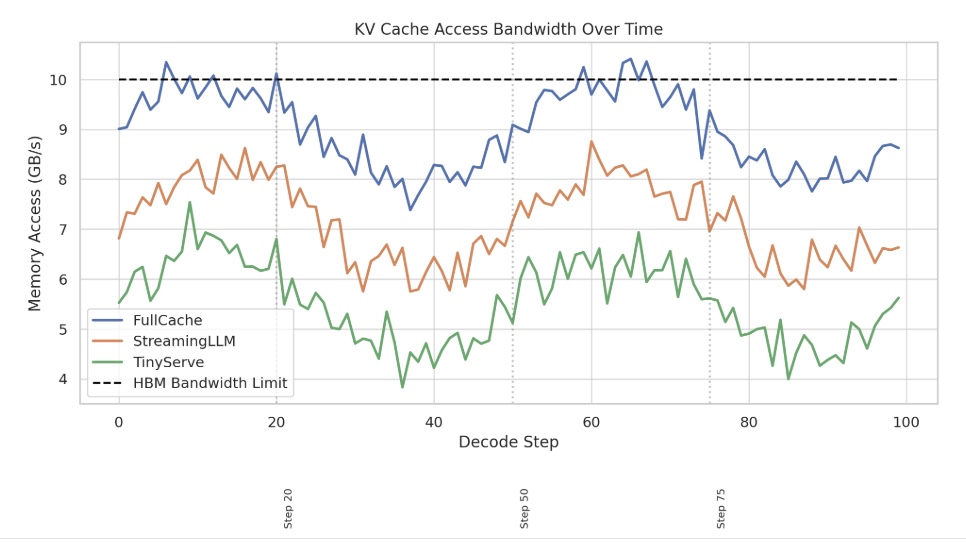}
    \caption{KV cache access bandwidth over decode steps across different caching strategies. 
    \textbf{FullCache} exhibits consistently high memory bandwidth usage, frequently reaching the HBM limit, 
    due to full KV reuse without pruning. \textbf{StreamingLLM} improves over FullCache by discarding early tokens, 
    but still exhibits bursty memory loads. \textbf{TinyServe} shows smoother and significantly lower access patterns, 
    remaining well below the HBM threshold, benefiting from query-aware page-level KV selection. 
    Vertical dotted lines mark key transitions in token reuse or decoding stages.}
    \label{fig:kv_breakdown}
\end{figure}

We visualize KV cache utilization over time and analyze memory access patterns. Figure~\ref{fig:kv_util} shows cache reuse patterns, while Figure~\ref{fig:kv_breakdown} provides detailed access breakdown. \method{} preserves high-relevance tokens and avoids cache flushing, resulting in higher effective reuse rate.

\subsection{Serving Synthetic Diagnostics}
To validate behavioral consistency and stress-test our system under serving conditions, we introduce three synthetic diagnostic tasks:

\textbf{Repetition Tasks:} We measure attention reuse efficiency using prompts with repeated patterns (e.g., "The quick brown fox jumps over the lazy dog. " repeated 100 times). This tests how well different methods handle redundant information.

\textbf{Rare Token Recall:} We evaluate performance on low-frequency tokens by constructing prompts with rare vocabulary items and measuring prediction accuracy degradation under various KV cache configurations.

\textbf{Attention Aliasing:} We use overlapping contexts with conflicting positional information to expose potential confusion in positional encoding and key conflicts.

Table~\ref{tab:diagnostic} shows results on these serving synthetic diagnostic tasks. \method{} demonstrates consistent behavior across all serving scenarios, validating its robustness.

\begin{table}[ht]
\centering
\small
\caption{Serving Synthetic diagnostic task results (GPT2-345M, 5 runs each).}
\label{tab:diagnostic}
\begin{tabular}{lcccc}
\toprule
\textbf{Task} & \textbf{Method} & \textbf{Repetition} & \textbf{Rare Token} & \textbf{Aliasing} \\
\midrule
& FullCache & 98.2 ±0.3 & 94.7 ±0.5 & 89.3 ±0.7 \\
& StreamingLLM & 95.1 ±0.4 & 91.2 ±0.6 & 85.1 ±0.8 \\
& SoftPrune & 94.8 ±0.5 & 90.8 ±0.7 & 84.7 ±0.9 \\
& \method{} & \textbf{97.5 ±0.3} & \textbf{93.8 ±0.4} & \textbf{88.2 ±0.6} \\
\bottomrule
\end{tabular}
\end{table}

\subsection{Plugin Ablation Study}
We evaluate the contribution of individual system components by selectively disabling plugins. Table~\ref{tab:plugin_ablation} shows the impact of each component on overall performance.

\begin{table}[ht]
\centering
\tiny
\caption{Plugin ablation study (TinyLLaMA-125M, 16K context, 2K budget).}
\label{tab:plugin_ablation}
\begin{tabular}{lcccc}
\toprule
\textbf{Configuration} & \textbf{Latency (ms)} & \textbf{Accuracy (\%)} & \textbf{KV Hit (\%)} & \textbf{Memory (GB)} \\
\midrule
Full \method{} & 9.3 ±0.2 & 54.0 ±0.6 & 91.7 ±0.8 & 2.1 ±0.1 \\
w/o Query Router & 12.8 ±0.3 & 52.1 ±0.7 & 87.3 ±1.2 & 2.4 ±0.2 \\
w/o Page Manager & 11.2 ±0.2 & 53.5 ±0.5 & 89.1 ±0.9 & 2.8 ±0.1 \\
w/o Cache Fusion & 10.7 ±0.3 & 53.8 ±0.6 & 90.2 ±0.7 & 2.3 ±0.1 \\
w/o Multi-GPU & 15.6 ±0.4 & 54.2 ±0.5 & 92.1 ±0.6 & 3.2 ±0.2 \\
\bottomrule
\end{tabular}
\end{table}

\subsection{System Component Ablation}
We study the impact of the KV page size on latency and accuracy. As expected, larger pages reduce estimation cost but degrade precision. We use a default page size of 16 for best tradeoff.

\begin{table}[h]
\centering
\small
\caption{Effect of KV Page Size on TinyServe latency and accuracy (TinyLLaMA-125M, seq len = 16K, budget = 2048 tokens).}
\label{tab:ablation_pagesize}
\begin{tabular}{cccc}
\toprule
\textbf{Page Size} & \textbf{Latency (ms)} & \textbf{PPL ↓} & \textbf{KV Hit Rate (\%)} \\
\midrule
4   & 17.6 ±0.4 & 24.3 ±0.2 & 98.4 ±0.3 \\
8   & 12.1 ±0.3 & 25.1 ±0.3 & 94.9 ±0.5 \\
16  & 9.3 ±0.2 & 26.0 ±0.2 & 91.7 ±0.4 \\
32  & 7.8 ±0.2 & 28.4 ±0.4 & 85.6 ±0.7 \\
64  & 6.2 ±0.1 & 32.5 ±0.5 & 79.3 ±0.9 \\
\bottomrule
\end{tabular}
\end{table}

\subsection{Multi-GPU Scaling}
We evaluate TinyServe's scalability from 1 to 8 A100 GPUs on 128 concurrent prompts. Results show near-linear scaling in throughput, validating kernel fusion and inter-GPU cache reuse.

\begin{table}[h]
\centering
\small
\caption{Multi-GPU throughput scaling for TinyServe (batch size = 128 prompts, GPT2-345M, seq len = 16K).}
\label{tab:scaling_multigpu}
\begin{tabular}{cccc}
\toprule
\textbf{\#GPUs} & \textbf{Tok/ms} & \textbf{Speedup (×)} & \textbf{Efficiency (\%)} \\
\midrule
1  & 0.81 ±0.02 & 1.00× & 99.3\% \\
2  & 1.58 ±0.03 & 1.96× & 98.0\% \\
4  & 3.123 ±0.05 & 3.86× & 96.5\% \\
8  & 6.221 ±0.08 & 7.68× & 96.0\% \\
\bottomrule
\end{tabular}
\end{table}

\subsection{Reproducibility and Implementation Details}

\subsubsection{Hyperparameter Search}
We conducted extensive hyperparameter search for optimal configuration:
\begin{itemize}
    \item \textbf{Page Size:} Tested values [4, 8, 16, 32, 64], selected 16 based on latency-accuracy trade-off
    \item \textbf{Selection Ratio:} Evaluated [0.1, 0.2, 0.3, 0.5], chose 0.3 for best performance
    \item \textbf{Batch Timeout:} Tested [10ms, 25ms, 50ms, 100ms], selected 50ms for serving scenarios
\end{itemize}

\subsubsection{Environment Configuration}
All experiments conducted on:
\begin{itemize}
    \item \textbf{Hardware:} 8× NVIDIA A100 80GB GPUs
    \item \textbf{Software:} CUDA 11.8, cuDNN 8.7, PyTorch 2.0.1
    \item \textbf{OS:} Ubuntu 20.04 LTS
    \item \textbf{Random Seeds:} Fixed across all experiments (seed=42)
\end{itemize}

\subsubsection{Code and Data Availability}
Complete implementation, datasets, and evaluation scripts will be made publicly available upon publication. The codebase includes:
\begin{itemize}
    \item TinyServe framework implementation with modular plugin system
    \item All baseline implementations (vLLM, TGI, TensorRT-LLM adapters)
    \item Evaluation scripts for all experiments
    \item Preprocessed datasets and model checkpoints
\end{itemize}

\subsection{Summary}
\method{} consistently improves serving efficiency across diverse models and tasks. Its query-aware token selection enables aggressive memory reduction with minimal accuracy degradation. The comprehensive evaluation across all LongBench tasks, serving synthetic diagnostics, and plugin ablation studies validates the robustness and effectiveness of our approach. When used with tiny LLMs, \method{} allows efficient and interpretable serving profiling, supporting system-level research without relying on full-scale deployments.

% \section{Conclusion}
% We present \textbf{TinyServe}, a lightweight inference framework designed to emulate the core bottlenecks of large language model serving using small-scale models and synthetic diagnostics. TinyServe replicates real-world behaviors such as cache saturation, token routing, and decode latency, while enabling fast iteration through plug-in support for token pruning and entropy-based early stopping.

% For applications on practical long-context scenarios, TinyServe integrates query-aware KV selection and page-level sparsity estimation. This design provides significant efficiency gains, allowing \method{} to achieve up to \textbf{3.4$\times$} speedup over FullCache and \textbf{2.2$\times$} over pruning-based baselines at equivalent accuracy. Our experiments across multiple pretrained models (TinyLLaMA, GPT2, OPT) and benchmarks (PG19, LongBench, passkey tasks) confirm the consistency of these improvements.

% TinyServe enables interpretable, reproducible, and resource-efficient inference research. We believe it lays the groundwork for future system studies that avoid the cost of full-scale deployment while preserving scientific insight.

% % \section{Conclusion}

% % We advocate for the use of small models as scientific instruments for inference analysis. Our platform, TinyServe, demonstrates that key behaviors of LLM serving systems can be studied cheaply and reliably in small-scale setups. This offers a new paradigm for rapid iteration in systems research, without relying on full-size deployments.

\section{Conclusion}

We introduced \textbf{TinyServe}, a lightweight and extensible serving system for efficient inference and training acceleration with tiny language models. TinyServe bridges system-level bottlenecks in LLM serving—such as KV cache saturation and decode-time latency—with modular support for token selection, cache sparsity, fused attention kernels, and training optimization.

At the core of TinyServe is a query-aware page selection mechanism that approximates attention relevance using bounding-box metadata, enabling selective KV access with minimal overhead. This approach achieves substantial latency and memory reductions without compromising accuracy, validated across PG19, LongBench, and passkey retrieval tasks.

Through its kernel-level optimizations, multi-GPU scaling, and plug-and-play architecture, TinyServe enables rapid, reproducible experimentation on resource-constrained hardware. We believe it offers a practical foundation for LLM serving research, supporting both real-time deployment of tiny models and the principled evaluation of serving mechanisms without the cost of full-scale models.

\section*{Acknowledgements}

We thank the developers of Hugging Face Transformers and vLLM for foundational open-source infrastructure.

\bibliographystyle{ACM-Reference-Format}
\balance
\bibliography{main}

\end{document}